\def\year{2020}\relax
\documentclass[letterpaper]{article} 
\usepackage{aaai20}  
\usepackage{times}  
\usepackage{helvet} 
\usepackage{courier}  
\usepackage[hyphens]{url}  
\usepackage{graphicx} 
\usepackage{graphicx}  
\usepackage{amsmath,amssymb}
\usepackage{comment}
\usepackage{graphicx}
\usepackage{booktabs}
\usepackage{graphicx}
\usepackage{bm}
\usepackage{ifthen}
\usepackage{enumitem}
\usepackage{algorithm}
\usepackage{algorithmic}
\usepackage{tabularx,array,booktabs}
\usepackage{lipsum}  
\usepackage{tikz} 
\usetikzlibrary{arrows,decorations.pathmorphing,backgrounds,fit,positioning,shapes.symbols,chains}
\tikzset{box1/.style={draw=black, thick, rectangle, minimum height=1.4cm, minimum width=3cm }}

\newcommand{\Section}[1]{\section{#1}\label{sec:#1}}
\newcommand{\Subsection}[1]{\subsection{#1}\label{sec:#1}}
\newcommand{\Subsubsection}[1]{\subsubsection{#1}\label{sec:#1}}

\title{Time your hedge with Deep Reinforcement Learning}
\author{
	Eric Benhamou \textsuperscript{\rm 1}, David Saltiel \textsuperscript{\rm 2,\rm 3}, Sandrine Ungari \textsuperscript{\rm 4}, Abhishek Mukhopadhyay \textsuperscript{\rm 5} \\
	\textsuperscript{\rm 1} MILES, LAMSADE, Dauphine university, France \texttt{eric.benhamou@lamsade.dauphine.fr}\\
	\textsuperscript{\rm 2} AI Square Connect, France \texttt{david.saltiel@aisquareconnect.com}\\
	\textsuperscript{\rm 3} LISIC, ULCO, France \\
	\textsuperscript{\rm 4} Societe Generale, Cross Asset Quantitative Research, UK,\\ 
	\textsuperscript{\rm 5} Societe Generale, Cross Asset Quantitative Research, France,\\ 
	\texttt{\{sandrine.ungari,abhishek.mukhopadhyay\}@sgcib.com}
}

\begin{document}

\maketitle
\thispagestyle{plain}
\pagestyle{plain}

\begin{abstract}
Can an asset manager plan the optimal timing for her/his hedging strategies given market conditions? 
The standard approach based on Markowitz or other more or less sophisticated financial rules aims to find the best portfolio allocation thanks to forecasted expected returns and risk but fails to fully relate market conditions to hedging strategies decision. In contrast, Deep Reinforcement Learning (DRL) can tackle this challenge by creating a dynamic dependency between market information and hedging strategies allocation decisions. In this paper, we present a realistic and augmented DRL framework that: (i) uses additional contextual information to decide an action, (ii) has a one period lag between observations and actions to account for one day lag turnover of common asset managers to rebalance their hedge, (iii) is fully tested in terms of stability and robustness thanks to a  repetitive train test method called anchored walk forward training, similar in spirit to k fold cross validation for time series and (iv) allows managing leverage of our hedging strategy. Our experiment for an augmented asset manager interested in sizing and timing his hedges shows that our approach achieves superior returns and lower risk.
\end{abstract}


\Section{Introduction}
From an external point of view, the asset management (buy side) industry is a well-suited industry to apply machine learning as large amount of data are available thanks to the revolution of electronic trading and the methodical collection of data by asset managers or their acquisition from data providers. In addition, machine based decision can help reducing emotional bias and taking rational and systematic investment choices \cite{Kahneman_2011}. However, to date, the buy side industry is still largely relying on old and traditional methods to make investment decisions and in particular to choose portfolio allocation and hedging strategies. It is hardly using machine learning in investment decisions. \\
This is in sharp contrast with the ubiquitous usage of deep reinforcement learning (DRL) in other industries and in particular its use for solving challenging tasks like autonomous driving \cite{Wang2018DeepRL}, learning advanced locomotion and manipulation skills from raw sensory inputs \cite{Levine_2015,Levine_2016,Schulman_2015,Schulman_2017,Lillicrap_2015} or on a more conceptual side for reaching supra human level in popular games like Atari \cite{mnih-atari-2013}, Go \cite{Silver_2016,silver2017mastering}, StarCraft II \cite{Vinyals_2019}, etc ...\\

It therefore makes sense to investigate if DRL can help help in financial planning and in particular in creating augmented asset managers. To narrow down our problem, we are specifically interested in finding hedging strategies for a risky asset. To make things concrete and more illustrative, we represent this risky asset in our experiment with the MSCI World index that captures large and mid cap securities across 23 developed financial markets. The targeted hedging strategies are on purpose different in nature and spirit. They are appropriate under distinctive market conditions. Financial planning is therefore critical for deciding the appropriate timing when to add and remove these hedging strategies. 

\Subsection{Related works}
At first, reinforcement learning was not used in portfolio allocation. Initial works focused on trying to make decisions using deep networks to forecast next period prices, \cite{Freitas_2009,Niaki2013,Heaton_2017}. Armed with the forecast, an augmented asset manager could solve its financial planning problem to decide the optimal portfolio allocations. However, this initial usage of machine learning contains multiple caveats. First, there is no guarantee that the forecast is reliable in the near future. On the contrary, it is a stylized fact that financial markets are non stationary and exhibit regime changes \cite{Salhi_2016,Dias_2015,Zheng_2019}, making the prediction exercise quite difficult and unreliable. Second, it does not target specifically the financial planning question of finding the optimal portfolio based on some reward metrics. Third, there is no consideration of online learning to adapt to changing environment as well as the incorporation of transaction costs.\\
A second stream of research around deep reinforcement learning has emerged to address these points \cite{Jiang_2016,Jiang_2017,Liang_2018,Yu_2019,Wang_2019,Liu_2020,Ye_2020,Li_2019,Xiong_2018,benhamou2020detecting}. The dynamic nature of reinforcement learning makes it an obvious candidate for changing environment  \cite{Jiang_2016,Jiang_2017,Liang_2018}. Transaction costs can be easily included in rules \cite{Liang_2018,Yu_2019,Wang_2019,Liu_2020,Ye_2020,Yu_2019}. However, these works, except \cite{Ye_2020} and \cite{benhamou2020detecting} rely only on time series of open high low close prices, which are known to be very noisy. Secondly, they all assume an immediate action after observing prices which is quite different from reality. Most asset managers need a one day turnaround to manage their new portfolio positions. Thirdly, except \cite{benhamou2020detecting}, they rely on a single reward function and do not measure the impact of the reward function. Last but not least, they only do one train and test period, never testing for model stability.

\Subsection{Contributions}
Our contributions are fourfold:
\begin{itemize}
\item \textbf{The addition of contextual information.} Using only past information is not sufficient to learn in a noisy and fast changing environment. The addition of contextual information improves results significantly. Technically, we create two sub-networks: one fed with direct observations (past prices and standard deviation) and another one with contextual information (level of risk aversion in financial markets, early warning indicators for future recession, corporate earnings etc...). 

\item \textbf{One day lag between price observation and action.} We assume that prices are observed at time $t$ but action only occurs at time $t+1$, to be consistent with reality. This one day lag makes the RL problem more realistic but also more challenging. 

\item \textbf{The walk-forward procedure.} Because of the non stationarity nature of time dependent data and especially financial data, it is crucial to test DRL models stability. We present a new methodology in DRL model evaluation referred to as walk forward analysis that iteratively trains and tests the model on extending data-set. This can be seen as the analogy of cross validation for time series. This allows validating that selected hyper parameters work well over time and that the resulting model is stable over time.

\item \textbf{Model leverage}. Not only do we do a multi inputs network, we also do a multi outputs network to compute at the same time the percentage in each hedging strategy and the overall leverage. This is a nice feature of this DRL model as it incorporates by design a leverage mechanism. To make sure the leverage is in line with the asset manager objective, we cap the leverage to the maximum authorized leverage, which is in our case 3. This byproduct of the method is another key difference with standard financial methods like Markwitz that do not care about leverage and only give a percentage for the hedging portfolio allocation.
\end{itemize}

\Section{Background and mathematical formulation}\label{sec:Background}
In standard reinforcement learning, models are based on Markov Decision Process (MDP) \cite{SuttonBarto_2018}.
A Markov decision process is defined as a tuple $\mathcal{M} = (\mathcal{X}, \mathcal{A}, p,r)$
where:
\begin{itemize}
\item $\mathcal{X}$ is the state space,
\item $\mathcal{A}$ is the action space,
\item $p(y|x, a)$ is the transition probability such that\\ $p(y|x, a) = \mathbb{P}(x_{t+1} = y|x_t = x, a_t = a)$,
\item $r(x, a, y)$ is the reward of transition (x, a, y).
\end{itemize}

MDP assumes that the we know all the states of the environment and have all the information to make the optimal decision in every state. The Markov property in addition implies that knowing the current state is sufficient.\\
From a practical standpoint, the general RL setting is modified by taking a pseudo state formed with a set of past observations $(o_{t-n}, o_{t-n-1}, \ldots, o_{t-1}, o_t)$. In practice to avoid large dimension and the curse of dimension, it is useful to reduce this set and take only a subset of these past observations with $j< n$ past observations, such that $0<i_1< \ldots < i_j$ and $i_k \in \mathbb{N}$ is an integer. The set $\delta_1 = (0,i_1, \ldots, i_j)$ is called the observation lags. In our experiment we typically use lag periods like (0, 1, 2, 3, 4, 20, 60) for daily data, where $(0,1,2,3,4)$ provides last week observation, $20$ is for the one-month ago observation (as there is approximately 20 business days in a month) and 60 the three-month ago observation.

\Subsection{Observations}
\Subsubsection{Regular observations}
There are two types of observations: regular and contextual information. Regular observations are data directly linked to the problem to solve. In the case of a trading framework, regular observations are past prices observed over a lag period $\delta = (0<i_1< \ldots < i_j)$. To renormalize data, we rather use past returns computed as  $r_t = \frac{p^k_t}{p^k_{t-1} } -1$ where $p^k_t$ is the price at time $t$ of the asset $k$. To give information about regime changes, our trading agent receives also empirical standard deviation computed over a sliding estimation window denoted by $d$ as follows $\sigma^k_t  = \sqrt{ \frac{1}{d} \sum_{u =t-d+1}^t \left( r_u - \mu \right)^2 }$, where the empirical mean $\mu$ is computed as $\mu = \frac{1}{d} \sum_{u =t-d+1}^t r_u$. Hence our regular observations is a three dimensional tensor $A_t = \left[ A^1_t, A^2_t\right]$
\begin{center}
with \,\, $A^1_t =  \left( \!
\begin{array}{c   }
r^1_{t-i_j} \,\,	... \,\, r^1_t \\
... \,\,... \,\, ...\\
r^m_{t-i_j} \,\,.... \,\, r^m_t
\end{array} \! \right)\! ,  \,\,
 A^2_t =  \left( \!
\begin{array}{c  }
\sigma^1_{t-i_j} 	\,\,	... \,\, \sigma^1_t\\
... \,\,... \,\, ...\\
\sigma^m_{t-i_j} \,\,.... \,\, \sigma^m_t
\end{array} \! \right)$
\end{center}
This setting with two layers (past returns and past volatilities) is quite different from the one presented in \cite{Jiang_2016,Jiang_2017,Liang_2018} that uses different layers representing closing, open high low prices. There are various remarks to be made. First, high low information does not make sense for portfolio strategies that are only evaluated daily, which is the case of all the funds. Secondly, open high low  prices tend to be highly correlated creating some noise in the inputs. Third, the concept of volatility is crucial to detect regime change and is surprisingly absent from these works as well as from other works like \cite{Yu_2019,Wang_2019,Liu_2020,Ye_2020,Li_2019,Xiong_2018}.

\Subsubsection{Context observation}\label{Context}
Contextual observations are additional information that provide intuition about current context. For our asset manager, they are other financial data not directly linked to its portfolio assumed to have some predictive power for portfolio assets. Contextual observations are stored in a 2D matrix denoted by $C_t$ with stacked past $p$ individual contextual observations. Among these observations, we have the maximum and minimum portfolio strategies return and the maximum portfolio strategies volatility. The latter information is like for regular observations motivated by the stylized fact that standard deviations are useful features to detect crisis. The contextual state writes as $
C^t =  \left( \!\!
\begin{array}{c  }
c^1_t 	\,\,	... \,\, c^1_{t-i_k}\\
... \,\,... \,\, ...\\
c^p_t 	\,\,.... \,\, c^p_{t-i_k}
\end{array} \!\!\! \right)$. The matrix nature of contextual states $C_t$ implies in particular that we will use 1D convolutions should we use convolutional layers. All in all, observations that are augmented observations, write as $O_t =[ A_t, C_t]$, with $A_t=[A^1_t, A^2_t]$ that will feed the two sub-networks of our global network.

\Subsection{Action}
In our deep reinforcement learning the augmented asset manager trading agent needs to decide at each period in which hedging strategy it invests. The augmented asset manager can invest in $l$ strategies that can be simple strategies or strategies that are also done by asset management agent. To cope with reality, the agent will only be able to act after one period. This is because asset managers have a one day turn around to change their position. We will see on experiments that this one day turnaround lag makes a big difference in results. As it has access to $l$ potential hedging strategies, the output is a $l$  dimension vector that provides how much it invest in each hedging strategy. For our deep network, this means that the last layer is a softmax layer to ensure that portfolio weights are between $0$ and $100\%$ and sum to $1$, denoted by $(p^1_t, ..., p^l_t)$. In addition, to include leverage, our deep network has a second output which is the overall leverage that is between 0 and a maximum leverage value (in our experiment 3), denoted by $lvg_t$. Hence the final allocation is given by $lvg_t \times (p^1_t, ..., p^l_t)$.

\Subsection{Reward}
There are multiple choices for our reward and it's a key point for the asset manager to decide the reward corresponding to his\ her risk profile.
\begin{itemize}
    \item A straightforward reward function is to compute the final net performance of the combination of our portfolio computed as the value of our portfolio at the last train date $t_T$ over the initial value of the portfolio $t_0$ minus one: $\frac{P_{t_T}}{P_{t_0}}-1$.
    \item Another natural reward function is to compute the Sharpe ratio. There are various ways to compute Sharpe ratio and we take explicitly the annualized Sharpe ratio. 
    This annualized Sharpe ratio computed from daily data is defined as the ratio of the annualized return over the annualized volatility $\mu / \sigma$. The intuition of the Sharpe ratio is to account for risk when comparing returns with risk is represented by volatility.
    \item The last reward we are interested in is the Sortino ratio.
    This metric is a variation of the Sharpe ratio where the risk is computed by the downside standard deviation whose definition is to compute the standard deviation only on negative daily returns $(\tilde{r}_t)_{t=0..T}$ . 
    Hence the downside standard deviation is computed by $\sqrt{250} \times  \text{StdDev}[ (\tilde{r}_t)_{t=0..T})]$.
\end{itemize}

\begin{figure}[!htbp]
\centering
\includegraphics[width= \linewidth]{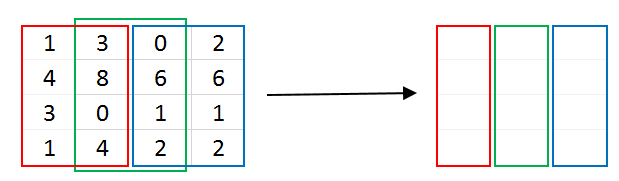}
\caption{2-D Convolution with stride of 1}\label{fig:conv1d}
\end{figure}

\Subsection{Convolutional network}
The similarities with image recognition (where pixels are stored in 3 different matrices representing red, green and blue image) enable us using convolution networks for our deep neural network. The analogy goes even further as it is well known in image recognition that convolutional networks achieve strong performances thanks to their capacity to extract meaningful features and to have very limited parameters hence avoiding over-fitting. Indeed, convolution allows us to extract features; blindly weighting locally the variables over the tensor. There is however something to notice. We use a convolution layer with a convolution window or kernel with a single row and a resulting vertical stride of 1. This particularity enables us to avoid mixing data from different strategies. We only mix data of the same strategies but for different observation dates. Recall that in convolution network, the stride parameter controls how the filter convolves around our input. Likewise the size of the window also referred to as the kernel size controls how the filter applies to data. Thus, a kernel with a row of 1 and a stride with a row of 1 allows us to detect the vertical (temporal) relation for each strategy by shifting one unit at a time, without mixing any data from different strategies. This concept is illustrated in figure \ref{fig:conv1d}.
Because of this peculiarity, we can interpret our 2-D convolution as an iteration over a 1-D convolution network for each variable.

\begin{figure}[!htbp]
\centering
\includegraphics[width= \linewidth]{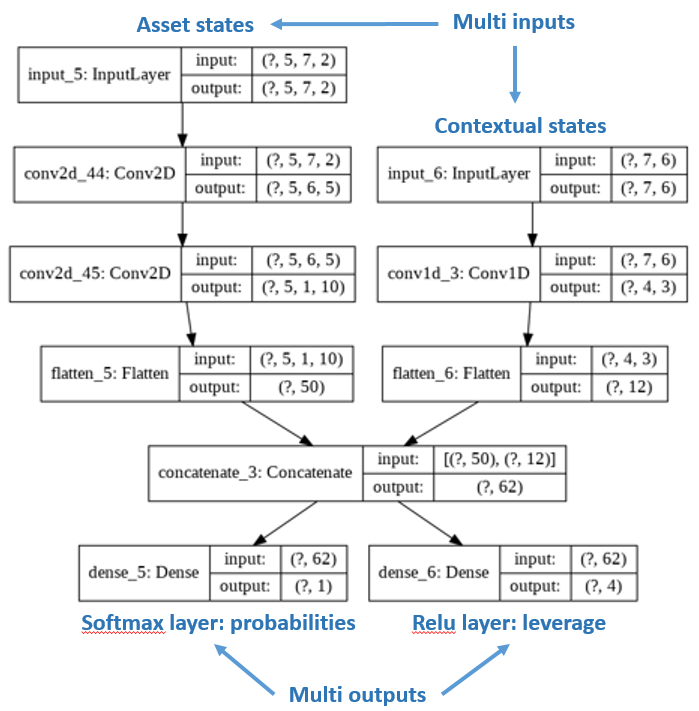}
\caption{network architecture obtained via tensorflow plotmodel function. Our network is very different from standard DRL networks that have single inputs and outputs. Contextual information introduces a second input while the leverage adds a second output}\label{fig:best_network}
\end{figure}

\Subsection{Multi inputs and outputs}
We display in figure \ref{fig:best_network} the architecture of our network. Because we feed our network with both data from the strategies to select but also contextual information, our network is a multiple inputs network.

Additionally, as we want from these inputs to provide not only percentage in the different hedging strategies (with a softmax activation of a dense layer) but also the overall leverage (with a dense layer with one single ouput neurons), we also have a multi outputs network. Additional hyperparameters that are used in the network as L2 regularization with a coefficient of 1e-8.

\Subsection{Adversarial Policy Gradient}
To learn the parameters of our network depicted in \ref{fig:best_network}, we use a modified policy gradient algorithm called adversarial as we introduce noise in the data as suggested in \cite{Liang_2018}.. The idea of introducing noise in the data is to have some randomness in each training to make it more robust. This is somehow similar to drop out in deep networks where we randomly pertubate the network by randomly removing some neurons to make it more robust and less prone to overfitting. 
A policy is a mapping from the observation space to the action space, $\pi:\mathcal{O}\rightarrow\mathcal{A}$. 
To achieve this, a policy is specified by a deep network with a set of parameters $\vec \theta$. The action is a vector function of the observation given the parameters: $\vec a_t = \pi_{\vec \theta}(\bm o_t)$.
The performance metric of $\pi_{\vec \theta}$ for time interval $[0,t]$ is defined as the corresponding total reward function of the interval $
	J_{[0,t]}(\pi_{\vec \theta}) = R\left( \vec o_1,\pi_{\vec \theta}(o_1),\cdots,
		\vec o_{t},\pi_{\vec \theta}(o_{t}),\vec o_{t+1} \right)
	\label{eq:policy_value}$.
After random initialization, the parameters are continuously updated along the gradient direction with a learning rate $\lambda$:
$\vec\theta \longrightarrow \vec\theta + \lambda\nabla_{\vec\theta}J_{[0,t]}(\pi_{\vec \theta})$. The gradient ascent optimization is done with standard Adam (short for Adaptive Moment Estimation) optimizer to have the benefit of adaptive gradient descent with root mean square propagation \cite{kingma2014method}. The whole process is summarized in algorithm \ref{alg1}.

\begin{algorithm}[!htbp]
    \caption{Adversarial Policy Gradient}
    \label{alg1}
\begin{algorithmic}[1]
    \STATE Input: initial policy parameters $\theta$, empty replay buffer $\mathcal{D}$
\REPEAT
    \STATE reset replay buffer
    \WHILE{not terminal}
        \STATE Observe observation $o$ and select action $a = \pi_{\theta}(o)$ with probability $p$ and random action with probability $1-p$, 
        \STATE Execute $a$ in the environment    
        \STATE Observe next observation $o'$, reward $r$, and done signal $d$ to indicate whether $o'$ is terminal
        \STATE apply noise to next observation $o'$
        \STATE store $(o,a,o')$ in replay buffer $\mathcal{D}$
        \IF{Terminal}
            \FOR{however many updates in $\mathcal{D}$}
                \STATE compute final reward $R$
            \ENDFOR
            \STATE update network parameter with Adam gradient ascent
                $\vec\theta \longrightarrow \vec\theta + \lambda\nabla_{\vec\theta}J_{[0,t]}(\pi_{\vec \theta})$
        \ENDIF
    \ENDWHILE
\UNTIL{convergence}
\end{algorithmic}
\end{algorithm}

In our gradient ascent, we use a learning rate of 0.01, an adversarial Gaussian noise with a standard deviation of 0.002. We do up to 500 maximum iterations with an early stop condition if on the train set, there is no improvement over the last 50 iterations.

\Subsection{Walk forward analysis}
In machine learning, the standard approach is to do $k$-fold cross validation as shown in figure \ref{fig:cross_val}. This approach breaks the chronology of data and potentially uses past data in the test set. Rather, we can take sliding test set and take past data as training data as show in the two sub-figures on the right of figure \ref{fig:walk_forward}. To ensure some stability, we favor to add incrementally new data in the training set, at each new step. This method is sometimes referred to as anchored walk forward as we have anchored training data. The negative effect of using extending training data set is to adapt slowly to new information. To our experience, because we do not have so much data to train our DRL model, we use anchored walk forward to make sure we have enough training data. Last but not least, as the test set is always after the train set, walk forward analysis gives less steps compared to cross validation. In practice for our data set, we train our models from 2000 to end of 2006 (to have at least seven years of data) and use a repetitive test period of one year.

\def\a{2}

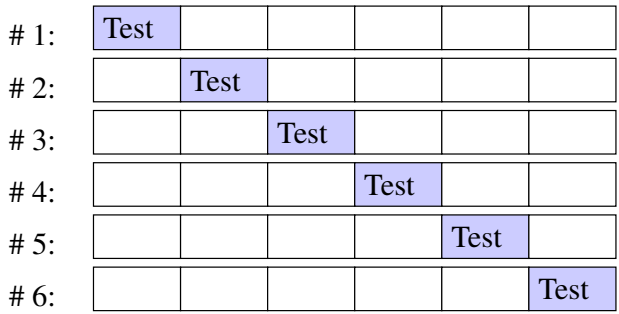
\begin{figure}[!htbp]
\centering
\resizebox{ \linewidth} {!} {
\begin{tikzpicture}[]
\foreach \i [count=\j from 0] in {white,white,white,white,white,blue!20} 
 \draw(\a+\j,0) rectangle (\a+\j+1,0.5)[fill=\i] ;
\foreach \i [count=\j from 0] in {white,white,white,white,blue!20,white} 
 \draw(\a+\j,0.6) rectangle (\a+\j+1,1.1)[fill=\i] ;
\foreach \i [count=\j from 0] in {white,white,white,blue!20,white,white} 
 \draw(\a+\j,1.2) rectangle (\a+\j+1,1.7)[fill=\i] ;
\foreach \i [count=\j from 0] in {white,white,blue!20,white,white,white} 
 \draw(\a+\j,1.8) rectangle (\a+\j+1,2.3)[fill=\i] ;
\foreach \i [count=\j from 0] in {white,blue!20,white,white,white,white} 
 \draw(\a+\j,2.4) rectangle (\a+\j+1,2.9)[fill=\i] ;
\foreach \i [count=\j from 0] in {blue!20,white,white,white,white,white} 
 \draw(\a+\j,3.0) rectangle (\a+\j+1,3.5)[fill=\i] ;
\foreach \i [count=\j from 0] in { , , , , ,} 
\draw(\a + 5.4-\j,\j*0.6) node[above]{Test};
\foreach \i [count=\j from 1] in { , , , , ,} 
\draw(\a-0.7,3.52-\j*0.6) node[above]{\# \j:};
\end{tikzpicture}
}
\caption{k-fold cross validation}
\label{fig:cross_val}
\end{figure}

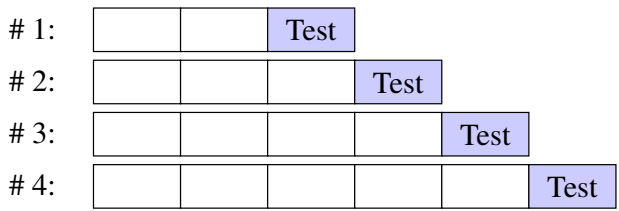
\begin{figure}[!htbp]
\centering
\resizebox{\linewidth} {!} {
\begin{tikzpicture}[]
\foreach \i [count=\j from 0] in {white, white, white,white,white,blue!20} 
 \draw(\a+\j,0) rectangle (\a+\j+1,0.5)[fill=\i] ;
\foreach \i [count=\j from 0] in {white, white,white,white,blue!20} 
 \draw(\a+\j,0.6) rectangle (\a+\j+1,1.1)[fill=\i] ;
\foreach \i [count=\j from 0] in {white, white,white,blue!20} 
 \draw(\a+\j,1.2) rectangle (\a+\j+1,1.7)[fill=\i] ;
\foreach \i [count=\j from 0] in {white, white,blue!20} 
 \draw(\a+\j,1.8) rectangle (\a+\j+1,2.3)[fill=\i] ;
\foreach \i [count=\j from -1] in { , , , } 
\draw(\a+4.5- \j,\j*0.6+0.6) node[above]{Test};
\foreach \i [count=\j from 1] in { , , , } 
\draw(\a-0.7,2.42-\j*0.6) node[above]{\# \j:};
\end{tikzpicture}
}
\caption{anchored walk forward}
\label{fig:walk_forward}
\end{figure}

\Section{Experiments}
\Subsection{Goal of the experiment}
We are interested in planing a hedging strategy for a risky asset. The experiment is using daily data from 01/05/2000 to 19/06/2020. The risky asset is the MSCI world index. We choose this index because it is a good proxy for a wide range of asset manager portfolios. The hedging strategies are 4 SG-CIB proprietary systematic strategies further described below .

\Subsection{Data-set description}\label{Data-set}
Systematic strategies are similar to asset managers that invest in financial markets according to an adaptive, pre-defined trading rule. Here, we use 4 SG CIB proprietary 'hedging strategies', that tend to perform when stock markets are down:
\begin{itemize}
\item Directional hedges - react to small negative return in equities,
\item Gap risk hedges - perform well in sudden market crashes,
\item Proxy hedges - tend to perform in some market configurations, like for example when highly indebted stocks under-perform other stocks,
\item Duration hedges - invest in bond market, a classical diversifier to equity risk in finance. 
\end{itemize}

The underlying financial instruments vary from put options, listed futures, single stocks, to government bonds. Some of those strategies are akin to an insurance contract and bear a negative cost over the long run. The challenge consists in balancing cost versus benefits.

In practice, asset managers have to decide how much of these hedging strategies are needed on top of an existing portfolio to achieve a better risk reward. The decision making process is often based on contextual information, such as the economic and geopolitical environment, the level of risk aversion among investors and other correlation regimes. The contextual information is modelled by a large range of features :

\begin{itemize}
    \item the level of risk aversion in financial markets, or market sentiment, measured as an indicator varying between 0 for maximum risk aversion and 1 for maximum risk appetite,
    \item the bond/equity historical correlation, a classical ex-post measure of the diversification benefits of a duration hedge, measured on a 1-month, 3-month and 1-year rolling window,
    \item The credit spreads of global corporate - investment grade, high yield, in Europe and in the US - known to be an early indicator of potential economic tensions, 
    \item The equity implied volatility, a measure if the 'fear factor' in financial market,
    \item The spread between the yield of Italian government bonds and the German government bond, a measure of potential tensions in the European Union,
    \item The US Treasury slope, a classical early indicator for US recession,
    \item And some more financial variables, often used as a gauge for global trade and activity: the dollar, the level of rates in the US, the estimated earnings per shares (EPS).
    
\end{itemize}
A cross validation step selects the most relevant features. In the present case, the first three features are selected. The rebalancing of strategies in the portfolio comes with transaction costs, that can be quite high since hedges use options. Transactions costs are like frictions in physical systems. They are taken into account dynamically to penalise solutions with a high turnover rate.

\Subsection{Evaluation metrics}
Asset managers use a wide range of metrics to evaluate the success of their investment decision. For a thorough review of those metrics, see for example \cite{Cogneau_2009}. The metrics we are interested in for our hedging problem are listed below:
\begin{itemize}
\item annualized return defined as the average annualized compounded return,
\item annualized daily based Sharpe ratio defined as the ratio of the annualized return over the annualized daily based volatility $\mu / \sigma$,
\item Sortino ratio computed as the ratio of the annualized return overt the downside standard deviation,
\item maximum drawdown (max DD) computed as the maximum of all daily drawdowns. The daily drawdown is computed as the ratio of the difference between the running maximum of the portfolio value ($RM_T = \max_{t=0..T}(P_t)$ ) and the portfolio value over the running maximum of the portfolio value. Hence $DD_T = (RM_T - P_T) / RM_T$ and $MDD_T = \max_{t=0..T}(DD_t)$. It is the maximum loss in return that an investor will incur if she/he invested at the worst time (at peak).
\end{itemize}

\Subsection{Baseline}

\Subsubsection{Pure risky asset}
This first evaluation is to compare our portfolio composed only of the risky asset (in our case, the MSCI world index) with the one augmented by the trading agent and composed of the risky asset and the hedging overlay. If our agent is successful in identifying good hedging strategies, it should improve the overall portfolio and have a better performance than the risky asset.

\Subsubsection{Markowitz}
In Markowitz theory \cite{Markowitz_1952}, risk is represented by the variance of the portfolio. Hence the Markowitz portfolio consists in maximizing the expected return for a given level of risk, represented by a given variance. Using dual optimization, this is also equivalent to minimize variance for a given expected return, which is solved by standard quadratic programming optimization. Recall that we have $l$ possible strategies and we want to find the best allocation according to the Sharpe ratio. Let $w = (w_1, ..., w_l)$ be the allocation weights with $1 \geq w_i \geq 0$ for $i=0 ... l$, which is summarized by $1 \geq w \geq 0 $, with the additional constraints that these weights sum to 1: $\sum_{i=1}^l w_i = 1$. \\
Let $\mu = ( \mu_1, ..., \mu_l)^T$ be the expected returns for our $l$ strategies and $\Sigma$ the matrix of variance covariances of the $l$ strategies' returns. Let $r_{min}$ be the minimum expected return. The Markowitz optimization problem to solve that is done by standard quadratic programming is the following:

\begin{eqnarray*}
&\text{Minimize} & w^T \Sigma w     \\
&\text{subject to}& \mu^T w \geq r_{min}  ,  \sum_{i=1 \ldots l} w_i = 1, w \geq 0 
\end{eqnarray*}

The Markowitz portfolio is a good benchmark very often used in portfolio theory as it allows investors to construct more efficient portfolios by controlling the variance of their strategies. One of the famous critic of this theory is that it controls the variance (and then the standard deviation) of the portfolio but it doesn't allow controlling a better risk indicator which is the downside standard deviation (representing the potential loss that may arise from risk compared to a minimum acceptable return).
Another limitation of this theory relies on the fact that it works under the assumption that investors are risk-averse. In other words, an investor prefers a portfolio with less risk for a given level of return and will only take on high-risk investments if he can expect a larger reward.

\Subsubsection{Follow the winner}
This is a simple strategy that consists in selecting the hedging strategy that was the best performer in the past year. If there is some persistence over time of the hedging strategies' performance, this simple methodology works well. It replicates standard investors behavior that tends to select strategies that performed well in the past.

\Subsubsection{Follow the loser}
As it name stands for, follow the loser is exactly the opposite of follow the winner. It assumes that there is some mean reversion in strategies' performance, meaning that strategies tend to perform equally well on long term and mean revert around their trend. Hence if a strategy did not perform well in the past, and if there is mean reversion, there is a lot of chance that this strategy will recover with its pairs.

\Subsection{Results and discussion}\label{sec:Results}
\begin{figure}[!htbp]
\centering
\includegraphics[width= 0.97 \linewidth]{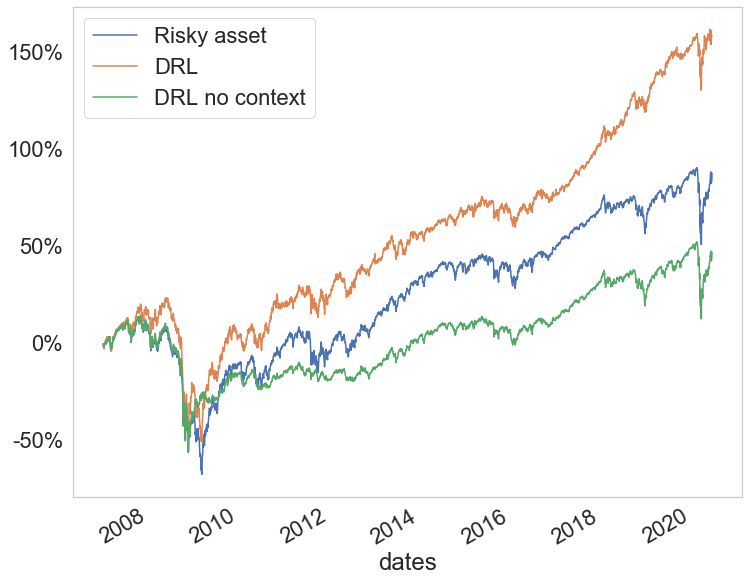}
\includegraphics[width= 0.97 \linewidth]{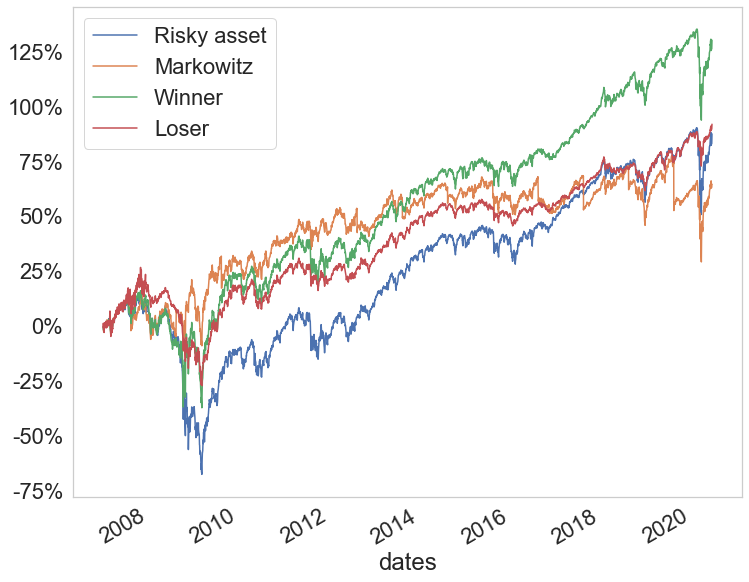}
\includegraphics[width= 0.97 \linewidth]{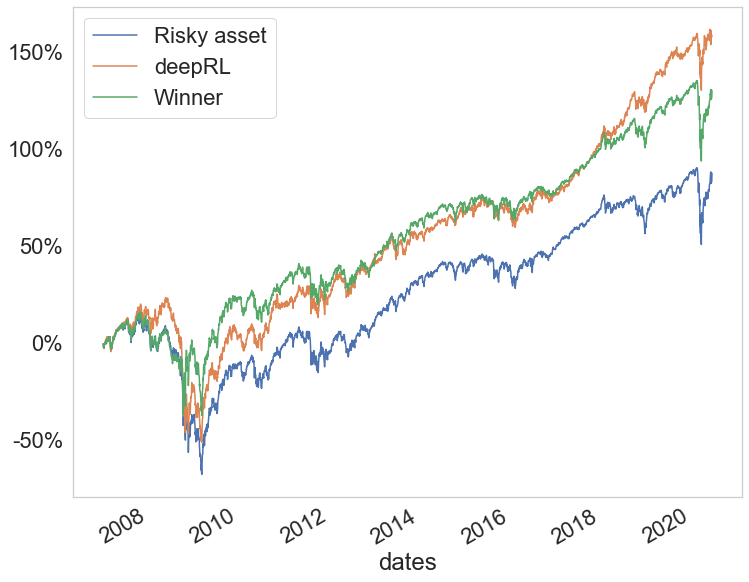}
\caption{performance of all models}\label{fig:all_models}
\end{figure}

\begin{figure}[!htbp]
\centering
\includegraphics[width= \linewidth]{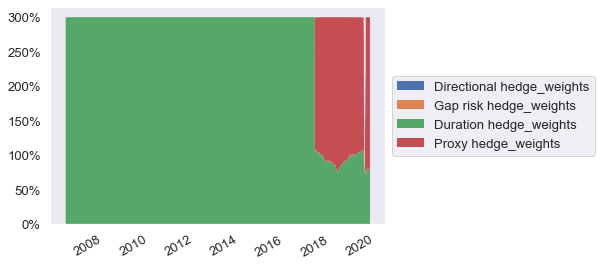}
\caption{DRL weights}\label{fig:drl_weights}
\end{figure}
\begin{figure}[!htbp]
\centering
\includegraphics[width= \linewidth]{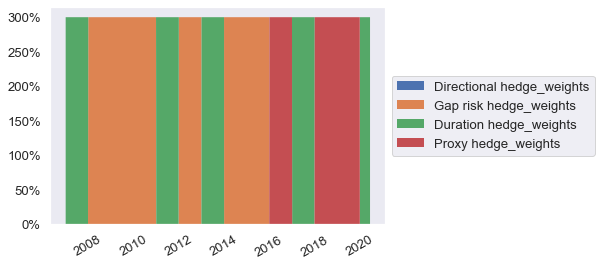}
\caption{Follow the winner weights}\label{fig:winner_weights}
\end{figure}
\begin{figure}[!htbp]
\centering
\includegraphics[width= \linewidth]{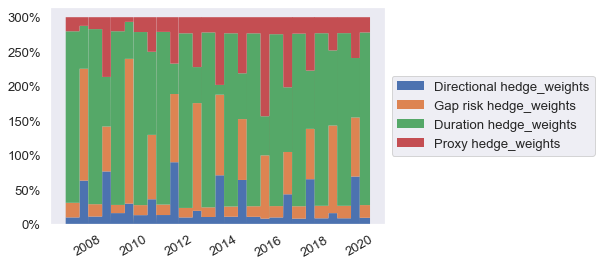}
\caption{Markowitz weights}\label{fig:markowitz_weights}
\end{figure}

\begin{table}[!htbp]
  \centering
  \caption{Models comparison over 3 and 5 years}\label{tab:Model comparison}
\resizebox{.5\textwidth} {!} {
    \begin{tabular}{|l|rrrr|}
    \toprule
 &        & 3 Years &   &\\
    \toprule   
        & \multicolumn{1}{l}{return } & \multicolumn{1}{l}{Sortino} & \multicolumn{1}{l}{Sharpe} & \multicolumn{1}{l|}{max DD}\\
    Risky asset & 10.27\% &       0.34  &      0.38  & -    0.34   \\
    DRL  & \textbf{22.45\%} & \textbf{1.18} & \textbf{1.17} & -0.27 \\
    Winner & 13.19\% & 0.66  & 0.72  & -0.35  \\
    Loser  & 9.30\% & 0.89  & 0.89  & \textbf{-0.15}    \\
    DRL no context & 8.11\% & 0.42  & 0.47  & -0.34   \\
    Markowitz & -0.31\% & -0.01  & -0.01  & -0.41   \\
    \toprule
     &        & 5 Years &   &\\
      \toprule   
        & \multicolumn{1}{l}{return } & \multicolumn{1}{l}{Sortino} & \multicolumn{1}{l}{Sharpe} & \multicolumn{1}{l|}{max DD}\\
    Risky asset & 9.16\% &       0.54  &       0.57  & -     0.34    \\
    DRL  & \textbf{16.42\%} & \textbf{0.98} & \textbf{0.96} & -0.27 \\
    Winner &  10.84\% & 0.65  & 0.68  & -0.35  \\
    Loser  &  7.04\% & 0.78  & 0.76  & \textbf{-0.15}    \\
    DRL no context &  6.87\% & 0.44  & 0.47  & -0.34   \\
    Markowitz & -0.07\% & -0.00  & -0.00  & -0.41  \\
    \bottomrule
    \end{tabular}
}
\end{table}%

We compare the performance of the following 5 models: DRL model based on convolutional networks with contextual states (Sentiment indicator, 6 month correlation between equity and bonds and credit main index), same DRL model without contextual states, follow the winner, follow the loser and Markowitz portfolio. The resulting graphics are displayed in figure \ref{fig:all_models} with the risky asset position alone in blue and the other models in orange, green and red. To make figures readable, we first show the two DRL models, with the risky asset and clearly see the impact of contextual information as the DRL model (in orange) is well above the green curve (the same model without contextual information) and is also well above the risky asset position alone (the blue curve). We then plot more traditional models like Markowitz, follow the Winner (entitled for space reason Winner) and follow the Loser (entitled for the same reason Loser). We finally plot the two best performers: the DRL and the Follow the Winner model, emphasizing that the difference between DRL and Follow the Winner is mostly in years 2018 to 2020 that exhibit regime changes, with in particular the recent Covid crisis. 

Out of these 5 models, only DRL and Follow the winner are able to provide significant net performance increase compared to the risky asset alone thanks to an efficient hedging strategy over the 2007 to 2020 period. The DRL model is in addition able to better adapt to the Covid crisis and to have better efficiency in net return but also Sharpe and Sortino ratios over 3 and 5 years as shown in table \ref{tab:Model comparison}. In addition, on the last graphic of figure \ref{fig:all_models}, we can remark that the DRL model has a tendency to move away from the blue curve (the risky asset) continuously and increasingly whereas the follow the winner model has moved away from the blue curve in 2015 and 2016 and tends to remain in parallel after this period, indicating that there is no continuous improvement of the model. The growing divergence of the DRL from the blue curve is a positive sign of its regular performance whci is illustrated in numbers in table \ref{tab:Model comparison}.

Moreover, when comparing the weights obtained by the different models (figures \ref{fig:drl_weights}, \ref{fig:winner_weights}, and \ref{fig:markowitz_weights}), we see that the bad performance of Markowitz can be a consequence of its diversification as it takes each year a non null position in the four hedging strategies and tends to change this allocation quite frequently. The rapid change of allocation is a sign of unstability of this method (which is a well known drawback of Markowitz). 

In contrast, DRL and Follow the winner models tends to choose only one or two strategies, in a stock picking manner. DRL model tends to choose mostly duration hedge and is able to dynamically adapt its behavior over the last 3 years and to better manage the Covid crisis with a mix allocation between duration and proxy hedge. 

In terms of the smallest maximum drawdown, the follow the loser model is able to significantly reduce maximum drawdown but at the price of a lower return, Sharpe and Sortino ratios. Removing contextual information deteriorates model performances significantly and is illustrated by the difference in term of return, Sharpe, Sortino ratio and maximum drawdown between the DRL and the DRL no context model. Last but not least, Markowitz model is not able to adapt to the new regime change of 2015 onwards despite its good performance from 2007 to 2015. It is the worst performer over the last 3 and 5 years because of this lack of adaptation. 

For all models, we use the walk forward analysis as described earlier. Hence, we start training the models from 2000 to end of 2006 and use the best model on the test set in 2007. We then train the model from 2000 to end of 2007 and use the best model on the test set in 2008 and etc ... In total, we do 14 training (from 2007 to 2020). This process ensures that we detect models that are unstable overtime and is similar in spirit to delayed online training.  We also provide in table \ref{tab:model choice} different configurations (adversarial training, use of context, and use of day lag), which leads to a total of 16 models. The frist 8 models are the ones with a daylag sorted in order of decreasing performance. The best model is the one with a reward in net profit, adversarial training, use of context information with a total performance of 81.8 \%. We also provide the corresponding same models but with no day lag (model 9 to 16). These models are theoreticall and not considered as they do not cope with reality.

\begin{table}[!htpb]
  \centering
  \caption{Model comparison based on reward function, adversarial training (noise in data) and use of contextual state}
  \label{tab:model choice}%
    \resizebox{0.48 \textwidth}{!}{
    \begin{tabular}{|l l c c c c |}
    \toprule
    \# &    reward  & adversarial? & context? & day lag & performance  \\
    \midrule
    \textbf{1} & \textbf{Net\_Profit}     &     \textbf{Yes}    &     \textbf{Yes}    &     \textbf{Yes}    &     \textbf{81.8\%} \\
    2 & net profit     & No     & Yes    & Yes    & 75.2\% \\
    3 & Sortino     & No     & Yes    & Yes    & 26.5\% \\
    4 & Sortino     & Yes    & Yes    & Yes    & 26.3\% \\
    5 & Sortino     & Yes    & No     & Yes    & -16.7\% \\
    6 & net profit     & Yes    & No     & Yes    & -29.5\% \\
    7 & Sortino     & No     & No     & Yes    & -45.0\% \\
    8 & net profit     & No     & No     & Yes    & -47.7\% \\
    \midrule
    \midrule
    9 & net profit     & Yes    & Yes    & No     & 193.8\% \\
    10 & net profit     & No     & Yes    & No     & 152.3\% \\
    11 & Sortino     & No     & Yes    & No     & 45.3\% \\
    12 & Sortino     & Yes    & Yes    & No     & 29.3\% \\
    13 & Sortino     & Yes    & No     & No     & 16.9\% \\
    14 & net profit     & Yes    & No     & No     & 13.9\% \\
    15 & Sortino     & No     & No     & No     & 10.6\% \\
    16 & net profit     & No     & No     & No     & 8.6\% \\
    \bottomrule
    \end{tabular}
    }
\end{table}

\Subsection{Impact of context}
In table \ref{tab:model choice}, we provide a list of 16 models based on the following choices: the choice of the reward function (net profit or Sortino), the use of adversarial training with noise in data or not, the use of contextual states, and the use of day lag between observations and actions. 

We see that the best DRL model with the day-lag turnover constraint is the one using convolutional networks, adversarial training, contextual states and net profit reward function. These 4 parameters are meaningful for our DRL model and change model performance substantially as illustrated by the table with a difference between model 1 (the best model) and model 8 (the worst model) of 129.5 \% (=81.8 \% - (-47.7 \%)). 

To measure the impact of the contextual information for our best model, we can measure it simply by doing the difference between model 1 and model 6 (as there are the same model except the presence or absence of contextual information). We find a significant impact as it accounts for 111.4 \% (=81.8 \% - (-29.5 \%)). It is quite intuitive that adding a context should improve the model as we provide more meaningful information to the model.

\Subsection{Impact of one day lag}
Our model accounts for the fact that asset managers cannot immediately change their position at the close of the financial markets. It is easy to measure the impact of the one day lag as we simply need to take the difference of performance between model 9 and model 1. We find an impact of the one day lag of 112 \% (= 193.8 \% - 81.8\%). This is like for contextual information substantial. It is not surprising that a delayed action (with one period lag) after observation makes the learning process more challenging for the DRL agent as influence of variables tends to decrease with time. Surprisingly, this salient modeling characteristic is ignored in existing literature \cite{Jiang_2017,Liang_2018,Yu_2019,Wang_2019,Liu_2020,Ye_2020,Li_2019}.

\Subsection{Future work}
As nice as this work is, there is room for improvement as we have only tested a few possible hyper-parameters for our convolutional networks and could play with more layers, other design choice like combination of max pooling layers (like in image recognition) and ways to create more predictive contextual information.

\Section{Conclusion}
In this paper, we address the challenging task of financial planning in a noisy and self adapting environment with sequential, non-stationary and non-homogeneous observations. Our approach is based on deep reinforcement learning using contextual information thanks to a second sub-network. We also show that the additional constraint of a delayed action following observations has a substantial impact that should not be overlooked. We introduce the novel concept of walk forward analysis to test the robustness of the deep RL model. This is very important for regime changing environments that cannot be evaluated with a simple train validation test procedure, neither a $k$-fold cross validation as it ignores the strong chronological feature of observations. 

For our trading agent, we take not only past performances of portfolio strategies over different rolling period, but also standard deviations to provide predictive variables for regime changes. Augmented states with contextual information make a big difference in the model and help the agent learning more efficiently in a noisy environment. On experiment, contextual based approach over-performs baseline methods like Markowitz or naive follow the winner and follow the loser. Last but not least, it is quite important to fine tune the numerous hyper-parameters of the contextual based DRL model, namely the various lags (lags period for the sub network fed by portfolio strategies past returns, lags period for common contextual features referred to as the common features in the paper), standard deviation period, learning rate, etc... 

Despite the efficiency of contextual based DRL models, there is room for improvement. Other information like news could be incorporated to continue increasing model performance. For large stocks, like tech stocks, sentiment information based on social media activity could also be relevant.

\subsubsection{Acknowledgments.}
We would like to thank Beatrice Guez and Marc Pantic for meaningful remarks while working on this project. The views contained in this document are those of the authors and do not necessarily reflect the ones of SG CIB.

\bibliography{main}
\bibliographystyle{aaai}
\end{document}